\documentclass[%
 reprint,
superscriptaddress,
 amsmath,amssymb,
 aps,
]{revtex4-2}

\usepackage{graphicx}
\usepackage{dcolumn}
\usepackage{bm}
\usepackage{acronym}
\usepackage{xcolor}
\usepackage{hyperref}
\hypersetup{
  colorlinks=true,
  citecolor=blue,   
  linkcolor=orange,
  urlcolor=blue
}

\begin{document}

\newacro{ADM}{Arnowitt--Deser--Misner}
\newacro{AMR}{adaptive mesh-refinement}
\newacro{BH}{black hole}
\newacro{BBH}{binary black-hole}
\newacro{BHNS}{black-hole neutron-star}
\newacro{BNS}{binary neutron star}
\newacro{CCSN}{core-collapse supernova}
\newacroplural{CCSN}[CCSNe]{core-collapse supernovae}
\newacro{CMA}{consistent multi-fluid advection}
\newacro{CFL}{Courant--Friedrichs--Lewy}
\newacro{DG}{discontinuous Galerkin}
\newacro{HMNS}{hypermassive neutron star}
\newacro{EM}{electromagnetic}
\newacro{ET}{Einstein Telescope}
\newacro{EOB}{effective-one-body}
\newacro{EOS}{equation of state}
\newacro{FF}{fitting factor}
\newacro{GR}{general-relativistic}
\newacro{GRLES}{general-relativistic large-eddy simulation}
\newacro{GRHD}{general-relativistic hydrodynamics}
\newacro{GRMHD}{general-relativistic magnetohydrodynamics}
\newacro{GW}{gravitational wave}
\newacro{KHI}{Kelvin--Helmholtz instability}
\newacro{ILES}{implicit large-eddy simulations}
\newacro{LIA}{linear interaction analysis}
\newacro{LES}{large-eddy simulation}
\newacroplural{LES}[LES]{large-eddy simulations}
\newacro{MHD}{ magnetohydrodynamics}
\newacro{MRI}{magnetorotational instability}
\newacro{NR}{numerical relativity}
\newacro{NS}{neutron star}
\newacro{PNS}{protoneutron star}
\newacro{RMNS}{remnant massive neutron star}
\newacro{SASI}{standing accretion shock instability}
\newacro{SGRB}{short $\gamma$-ray burst}
\newacro{SPH}{smoothed particle hydrodynamics}
\newacro{SN}{supernova}
\newacroplural{SN}[SNe]{supernovae}
\newacro{SNR}{signal-to-noise ratio}
\newacro{ZAMS}{zero age main sequence}

\preprint{APS/123-QED}

\title{Turbulent Dynamo Action in Binary Neutron Star Mergers}

\author{Eduardo M. Guti\'errez}
\email{emgutierrez@psu.edu}
 \affiliation{Institute for Gravitation and the Cosmos, The Pennsylvania State University, University Park, PA 16802, USA}
\affiliation{Department of Physics, The Pennsylvania State University, University Park, PA 16802, USA}

\author{David Radice}%
\affiliation{Institute for Gravitation and the Cosmos, The Pennsylvania State University, University Park, PA 16802, USA}
\affiliation{Department of Physics, The Pennsylvania State University, University Park, PA 16802, USA}
\affiliation{Department of Astronomy \& Astrophysics, The Pennsylvania State University, University Park, PA 16802, USA}

\author{Jacob Fields}
\affiliation{Institute for Advanced Study, Princeton, NJ 08540, USA}

\author{James M. Stone}
\affiliation{Institute for Advanced Study, Princeton, NJ 08540, USA}

\date{\today}

\begin{abstract}
We perform general-relativistic magnetohydrodynamics zoom-in simulations of the merger of two neutron stars starting with pulsar-like realistic initial magnetic field. We employ unprecedented grid resolution down to $\Delta x \simeq 2.9$~m. We find rapid amplification of the magnetic energy at small scales, and demonstrate the action of a small-scale turbulent dynamo. We find evidence that ordered large-scale magnetic structures emerge already during the early post-merger evolution. These results indicate that strong magnetic fields can be established promptly after merger, with important consequences for the electromagnetic counterparts of binary neutron star mergers.
\end{abstract}

\maketitle


\noindent \paragraph*{Introduction.}
\label{sec:intro}

---Binary neutron star mergers are among the strongest sources of gravitational waves~\cite{Cutler:1992tc, LIGOScientific:2017vwq} and are the only confirmed astrophysical site of heavy $r$-process elements~\cite{Kasen:2017sxr, Metzger:2019zeh, Watson:2019xjv}.
They also power bright electromagnetic transients, including kilonovae~\cite{LIGOScientific:2017ync, Arcavi:2017xiz, Smartt:2017fuw, Evans:2017mmy, Coulter:2017wya, Kasliwal:2017ngb} and gamma-ray bursts~\cite{LIGOScientific:2017zic, Ghirlanda:2018uyx, Mooley:2018qfh, Gutierrez:2024pch}.

Binary \ac{NS} systems merge at the end of a ${\sim}10{-}10^4$~Myr \ac{GW} driven inspiral \cite{Postnov:2014tza}.
This timescale is comparable to, or potentially larger than, the inferred decay time for pulsar magnetic fields \cite{1992ApJ...395..250G, Pons:2019zyc}.
As such, the \acp{NS} are not expected to be strongly magnetized at the time of merger.
However, strong, ordered magnetic fields are necessary to produce relativistic outflows matching the energies inferred from gamma-ray bursts.
This is exemplified by GW170817, the first \ac{NS} merger to be observed in both gravitational and electromagnetic waves.
This system is inferred to have merged $\gtrsim 1\ {\rm Gyr}$ after the birth of the second \ac{NS} \cite{LIGOScientific:2017apx, Levan:2017ubn}, but nevertheless produced a strong relativistic jet that was observed slightly off-axis \cite{Ghirlanda:2018uyx, Mooley:2018qfh}.
The presence of strong electromagnetic fields also provides a natural explanation for the UV/optical light curves on the first day of the merger, the so-called blue kilonova \cite{Metzger:2018qfl, Mosta:2020hlh, Combi:2023yav}.
Turbulence during the merger must have amplified the initial magnetic field, but the details of this process are far from understood~\cite{Radice:2024gic}.

In a seminal paper, Price and Rosswog \cite{Price:2006fi} first demonstrated, with Newtonian smooth particle hydrodynamics simulations, that the \ac{KHI} in the shear layer formed when the stars first come into contact can strongly amplify the initial magnetic field of the binary.
These results were later confirmed by both local and global grid-based simulations with adaptive mesh refinement \cite{Obergaulinger:2010gf, Zrake:2013mra, Giacomazzo:2014qba, Kiuchi:2014hja, Kiuchi:2015sga, Chabanov:2022twz}.
However, to power collimated outflows, the magnetic field in the remnant does not only need to be strong, but also coherent at large scales \cite{Narayan:2003by}.
Tangled magnetic fields produced by the \ac{KHI} are expected to undergo a reverse cascade and produce large-scale fields on a timescale of the order of
the Alfv\'en time $t_A \gtrsim 1$~second.
Numerical evidence that this phenomenon takes place after merger has been reported by Aguilera-Miret and collaborators on the basis of large-eddy simulations with subgrid models \cite{Aguilera-Miret:2023qih, Aguilera-Miret:2025nts}.
However, this process is too slow to play a dynamical role before black hole formation for most systems.
In a parallel line of investigation, Kiuchi and collaborators have shown that large-scale fields, if they exist in the first place, can be self-consistently
supported by dynamo action in the shear layers of the remnant after merger \cite{Kiuchi:2023obe}.
Their work, as well as most of the literature, considers binaries that already have strong ordered fields before merger in anticipation of the fact that such fields will be generated by the \ac{KHI} \cite{Combi:2023yav, Musolino:2024sju}.
However, the ability of the \ac{KHI} to generate the required large-scale magnetic fields in the few milliseconds it operates has not been previously demonstrated.
In this work, we present strong positive evidence supporting this idea.

\paragraph*{Methods.}
---We perform general-relativistic \ac{MHD} simulations with \texttt{AthenaK}~\cite{2024arXiv240916053S,Zhu:2024utz,Fields:2024pob} of an equal mass ($M_1=M_2=1.3\ M_\odot$) binary \ac{NS} system merging at the end of a gravitational-wave driven inspiral.
We initialize the magnetic field assuming each star has a purely dipolar field, with the dipole aligned with the binary angular momentum vector.
The maximum field, achieved at the center of the stars, is ${\simeq} 2 \times 10^{12}$~G, corresponding to a realistic maximum surface field of ${\simeq} 3\times 10^{10}~{\rm G}$.
As discussed in more detail in the Supplemental Material~\cite{SupplementalMaterial}, we first perform a global, adaptive mesh refinement simulation of the binary, with maximum resolution of $2^{-4} \ G M_\odot/c^2 \simeq 92\ {\rm m}$.
Then, we follow this up with a set of ``zoom-in'' simulations, achieving much higher resolution.
The zoom-in simulations consider a cubical region $[-L, L]^3$ of side length $2L$, with $L=4\ G M_\odot/c^2 \simeq 5.91\ {\rm km}$ centered at the (Newtonian) center of mass of the system.
Initial and boundary data, as well as the time-dependent background spacetime geometry, are obtained from the global \ac{NS} merger simulation.

This mapping allowed us to achieve unprecedented grid resolutions of down to $\Delta x = 2^{-9} \ G M_\odot/c^2 \simeq 2.9\ {\rm m}$.
Simulations with $\Delta x = 2^{-\ell}\ G M_\odot/c^2$ for $\ell = 4,5,6,7,8$ are carried out in an unigrid setup, while the highest resolution calculation uses a static mesh refinement grid with $\Delta x = 2^{-9}\ G M_\odot/c^2$ in the inner $[-L/4, L/4]^3$, and by a factor of two in the surrounding region $[-L/2, L/2]^3$ and four in the remainder of the grid (outer layers).
The combination of unigrid and mesh-refinement calculations allows us to quantify the importance of the non-local nonlinear dynamics.
Our simulations do not include explicit viscosity or resistivity; however, the width of the shear layer is limited by the scale height on the surface of the \acp{NS} ($\rho/\partial_r \rho$), which is of the order of $50$~m.
In other words, gravity introduces a length scale in this system, so initially the flow is not expected to exhibit self-similar behavior to arbitrarily small scales.

\paragraph*{Field Amplification.}
\begin{figure*}
    \centering
    \includegraphics[width=0.95\linewidth]{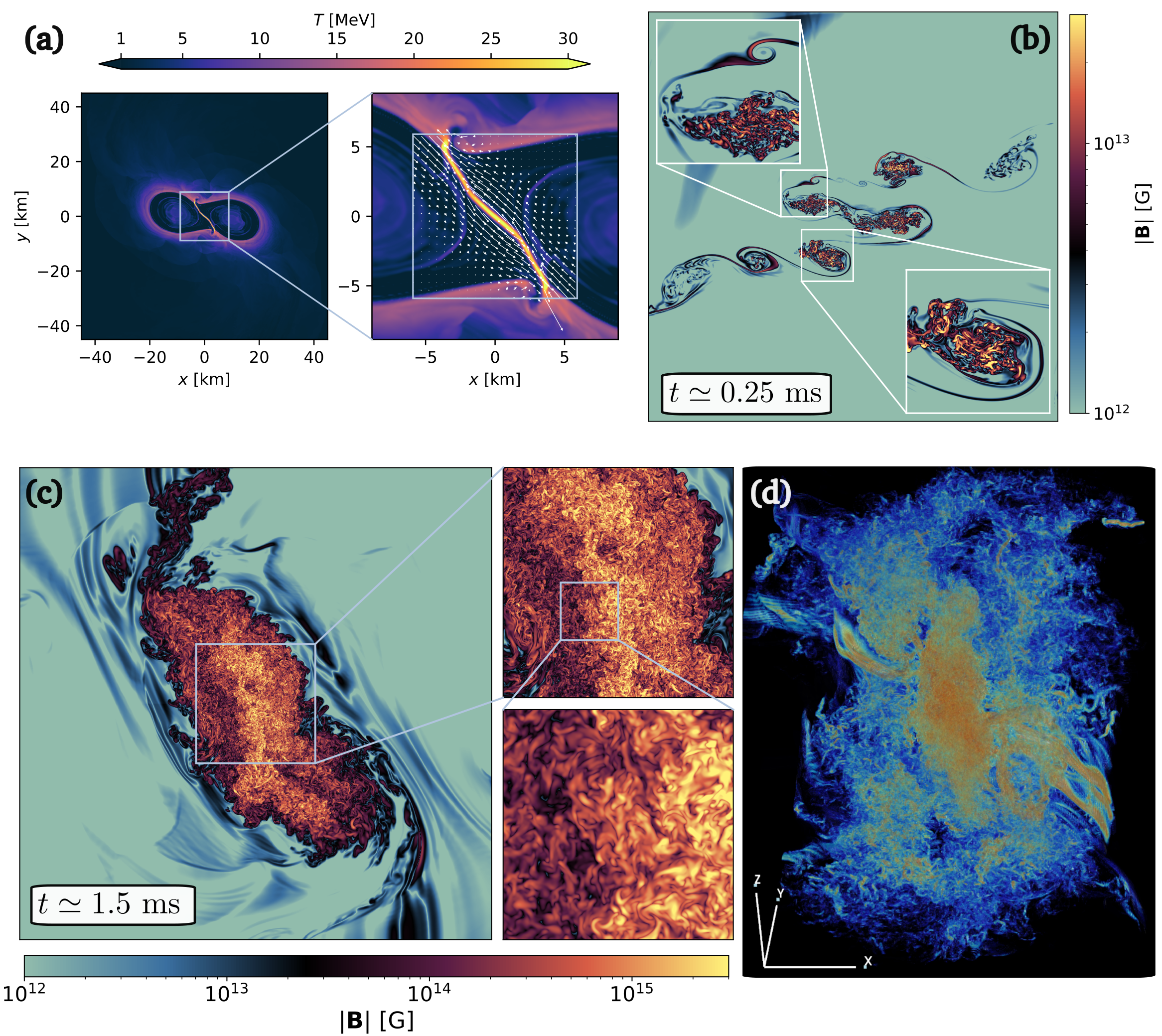}
    \caption{
    {\bf Turbulent magnetic-field amplification in the merger shear layer.}
    {\bf (a)} Equatorial snapshot of temperature at the time of hand-off; the right subpanel shows a zoom-in of the hot shear layer region with the arrows representing the velocity vector field in the co-rotating frame.
    The white box on the right subplot shows the box used in the zoom-in simulation.
    {\bf (b)} Magnetic field strength at $ t{\simeq} 0.25$ ms in the zoom-in simulation.
    The shear layer has rotated approximately $90^\circ$ and vortices of different sizes have already formed along it due to the development of the KHI; in these vortices, the magnetic field is amplified.
    {\bf (c)} Magnetic field strength at $t{\simeq} 1.5 ~{\rm ms}$.
    The vortices have merged, and turbulence is fully developed in a thick rotating layer.
    Note that the upper limit for the magnetic field colormap is higher than in {\bf (b)}.
    {\bf (d)} Three-dimensional rendering of the magnetic field strength at $t\approx 0.6~{\rm ms}$, showing the global structure of the field.}
    \label{fig:composition}
\end{figure*}

---The qualitative dynamics of the simulation and our workflow are summarized in Fig.~\ref{fig:composition}.
At the time of merger, the binary develops a very hot shear layer with a velocity jump of $\Delta v \approx 0.3c$.
This layer rapidly generates vortical structures where the field undergoes exponential amplification.
Initially, such amplification is confined to a narrow region at the interface between the stars.
At ${\approx} 1.5~{\rm ms}$, the shear layer grows in size as vortices produced by the primary \ac{KHI} themselves become KH-unstable (panel (b)).
Within approximately one rotational period $\tau \sim 1\ {\rm ms}$, all the initial unstable vortices have merged, and a large fraction of the box volume is filled with turbulent, vortical structures with a strongly amplified field (panel (c)).
The turbulent layer is not confined to the equatorial plane but also extends vertically, as shown in the lower-right panel, which displays a 3D rendering of the magnetic field strength.

\begin{figure}
    \centering
    \includegraphics[width=0.4\textwidth]{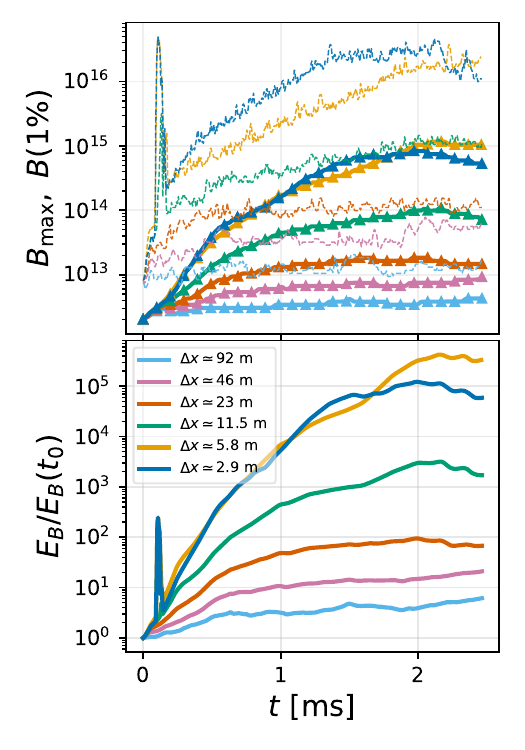}
    \caption{
    {\bf Magnetic field amplification in the early post-merger phase.} 
    {\em Upper panel:} 99th mass-weighted percentile of the magnetic field strength (thick lines with triangular markers), meaning that only 1\% of the mass of the fluid mass attains larger values, together with the maximum field strength (dashed lines).
    {Lower panel:} Time evolution of the total magnetic energy within the simulation domain, normalized to its initial value.}
    \label{fig:timeseries}
\end{figure}

Magnetic field amplification is highly intermittent: the field is tightly wound up in small-scale turbulent eddies, where it is locally amplified by orders of magnitude---reaching strengths of ${\sim}\textrm{few} \times 10^{16}\ {\rm G}$---before undergoing turbulent reconnection \cite{Lazarian_etal2020PhPl...27a2305L} (see Fig.~\ref{fig:timeseries}).
This degree of intermittency can be quantified in terms of the cumulative distribution function for the magnetic field \cite{Charbonneau_etal2001SoPh..203..321C}, which we find to form a clear power-law tail $p(B>B_0) \sim B_0^{-p}$ with $p \lesssim 0.2$ (see Supplemental Material~\cite{SupplementalMaterial}).

The total magnetic field energy (Fig.~\ref{fig:timeseries}) appears to saturate within ${\sim}2\ {\rm ms}$, a timescale that is comparable to the rotational frequency of the system.
This is long compared to the eddy turnover timescale of the region we are considering, $\tau_E \sim L/(0.1 c)\simeq 0.2\ {\rm ms}$, or the even shorter eddy turnover time for the vortical structures seeded by the \ac{KHI}.
We do not interpret this behavior as evidence of true physical saturation, but rather as the combination of two factors.
First, the field achieves a critical balance between turbulent amplification and numerical dissipation in the core of the most intense vortical structures (Fig.~\ref{fig:timeseries}).
This balance depends on the grid resolution employed in the simulations because of the lack of explicit resistivity in the simulations.
Second, it is the growth in size of the turbulent region that controls the overall energy budget for the remnant.
This interplay of very small and large-scale dynamics is a hallmark of MHD turbulence.

The integrated quantities in Fig.~\ref{fig:timeseries} seem to achieve convergence at resolutions finer than $\sim\! 5.8~{\rm m}$. 
However, this is likely a consequence of the fact that the finest refinement level occupies only a small fraction of the total simulation volume, and therefore should not be interpreted as true convergence.
In particular, volume-integrated quantities can depend on the size of the zoom-in region and on how much of the total volume is covered by the finest grid.
By contrast, pointwise maximum magnetic field strengths continue to increase with resolution, showing no signs of saturation.


\paragraph*{Dynamo Action.}
\begin{figure*}
    \centering
    \includegraphics[width=0.9\textwidth]{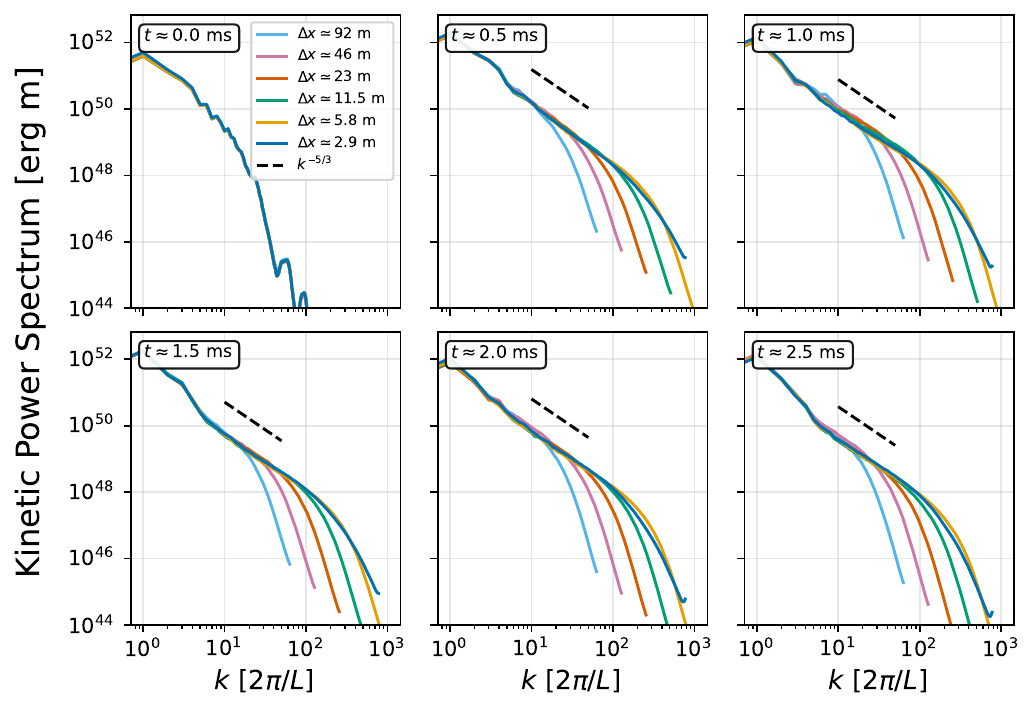}
    \caption{{\bf Kinetic energy power spectra at selected times for the six simulations.} The short black dashed line shows the Kolmogorov scaling $\sim\! k^{-5/3}$; times and numerical resolutions are indicated in the in-panel labels.}
    \label{fig:spectra_v}
\end{figure*}

\begin{figure*}
    \centering
    \includegraphics[width=0.9\textwidth]{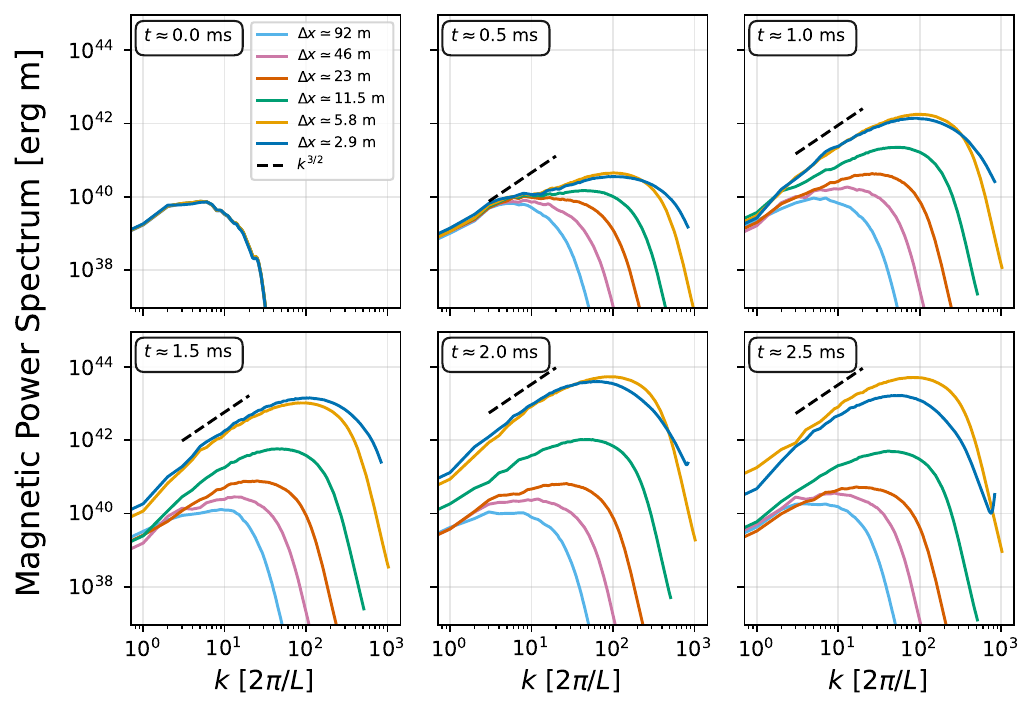}
    \caption{{\bf Magnetic energy spectra at selected times for the six simulations.} The short black dashed line shows the Kazantsev scaling $\sim\! k^{3/2}$; times and numerical resolutions are indicated in the in-panel labels.}
    \label{fig:spectra_b}
\end{figure*}

---We investigate the nature of turbulence in our simulations by calculating the power spectra for the density-weighted Eulerian-frame velocity $\sqrt{\rho}\, \mathbf{v}$ and magnetic field $\mathbf{B}$, shown in Figs.~\ref{fig:spectra_v} and~\ref{fig:spectra_b}.
Both velocity and magnetic spectra are fully developed after approximately one turnover time.
The velocity power spectrum shows a clear Kolmogorov $k^{-5/3}$ scaling for $k \gtrsim 10$, indicative of a fully developed turbulent flow.
The kinetic inertial range extends over more than a decade in $k$ space in the highest resolution simulations.
The magnetic energy spectrum quickly develops an extended region with power-law behavior with scaling $k^p$, with $p \simeq 0.75$, somewhat shallower than $k^{3/2}$ expected in a Kazantsev turbulent dynamo.
The spectrum peaks at increasingly small scales for higher resolution simulations, implying that the structure of the field is increasingly more tangled as the turbulence is better resolved.
In the earliest panel, the spectral peak reflects the characteristic scale of the seeded pulsar-like magnetic-field configuration rather than the turbulent dynamo.
We also note that at $t = 2.5$ ms the spectral magnetic power of the $\Delta x = 5.8$ m unigrid run exceeds that of the $\Delta x = 2.9$ m run because the power spectrum has contributions from the outer, lower-resolution volume in the static-mesh refinement setup of the latter.
Equipartition between magnetic and kinetic energy is not achieved in our simulations (see Supplemental Material~\cite{SupplementalMaterial}).

At late times, in the highest-resolution simulations, the dynamics change qualitatively, and the magnetic field grows also at large scales.
We interpret this as a consequence of the Kazantsev dynamo: its growing eigenfunctions peak near the resistive dissipation scale but carry a tail extending to large scales, which only becomes active once the inertial range is sufficiently broad. This interpretation is consistent with the simulations remaining in the kinematic regime throughout the simulation, as the magnetic energy stays well below equipartition with the kinetic energy.

This is also reflected by the steepening of the magnetic spectrum to $k^{3/2}$ and its growth at the largest scales, and is also consistent with the growth of the low-order magnetic multipoles shown in the Supplemental Material~\cite{SupplementalMaterial}.
Moreover, the fact that the growth in the large scale field components is seen in the highest resolution simulations, and not in the lower resolution calculations, which use the same initial and boundary conditions, demonstrates that this effect is not due to the large-scale dynamics of the system (e.g., due to the advection of magnetic flux into the simulation volume), but rather the result of the intrinsic turbulent dynamo dynamics.

\paragraph*{Discussion.}
---Our simulations show that the \ac{KHI} in the context of binary \ac{NS} mergers can operate as a turbulent dynamo, creating magnetar-level fields within milliseconds of contact.
Although the field remains dominated by small-scale structure, we find substantial amplification across all scales, indicating that a significant large-scale component emerges self-consistently from the resolved dynamics.

The resulting field strengths are sufficient to leave an imprint in the post-merger gravitational-wave signal \cite{Tsokaros:2024wgb, Rainho:2025ykl}, whose measurement is a key target for next-generation ground-based detectors such as the NEMO \cite{Ackley:2020atn}, the Einstein Telescope \cite{Punturo:2010zz}, and Cosmic Explorer \cite{Reitze:2019iox}.
If such fields were to emerge from the remnant surface, they could power gamma-ray flares \cite{Most:2023sft} and potentially explain recently observed quasi-periodic oscillations (QPOs) in some short gamma-ray bursts \cite{Chirenti:2023dzl}.
These QPOs have been interpreted as oscillation modes of the remnant \cite{Guedes:2024zkh}; however, within the standard paradigm for neutron-star merger dynamics, such oscillations are expected to be rapidly damped by gravitational-wave radiation reaction \cite{Bernuzzi:2015opx, Zappa:2017xba}.
Whether the field strengths achieved in our simulations are sufficient to undergo buoyant breakout remains an open question \cite{Fields_etal2026PhRvD.113d3056F}, and further amplification may be required for this mechanism to operate.

Overall, our results show that strong magnetic fields with a non-negligible large-scale component are present immediately after the merger.
If breakout occurs, these fields could launch relativistic jets before the remnant oscillations are fully damped.
They could also centrifugally accelerate the merger debris to trans-relativistic velocities, which would produce an UV/optical flash minutes to hours after merger \cite{Metzger:2018qfl, Combi:2023yav}, whose presence could be tested with future observations.

\noindent
\paragraph*{Acknowledgments.}

EG and DR were supported by the National Science Foundation under Grant PHY-2407681.
DR also acknowledges support from the Sloan Foundation, from the Department of Energy, Office of Science, Division of Nuclear Physics, under Award Number
DE-SC0024388, from the National Science Foundation under Grants No. PHY-2020275, PHY-2116686, and PHY-2512802.
DR, JF, and JS acknowledge support from the National Aeronautics and Space Administration under Grant No. 80NSSC25K213.
Zoom-in simulations were performed on Aurora at the Argonne Leadership Computing Facility, which is a DOE Office of Science User Facility supported under Contract DE-AC02-06CH11357. An award of computer time was provided by the INCITE program. Global simulations were performed on Perlmutter, at the National Energy Research Scientific Computing Center (NERSC), a Department of Energy User Facility, using NERSC award ERCAP0031370.
Large-language models (ChatGPT and Claude) were used to aid in coding some of the analysis scripts.

\nocite{1988CF.....16..257E,2008JCoPh.227.3191B,2010JGRA..11510112W,2012LNP...846.....G,2020ApJS..249....4S,2025MNRAS.542.2669B,2025NatAs...9.1195B,Anton:2005gi,Banyuls:1997zz,Barrera_1985,Daszuta:2021ecf,Duez:2008rb,Einfeldt1991,Espino:2023dei,Gardiner:2005hy,Gardiner:2007nc,Gottlieb2001,Gourgoulhon:2000nn,Harten1983,Hilditch:2012fp,Lorimer2008LRR....11....8L,Perego:2019adq,Steiner:2012rk}

\bibliography{references}

\clearpage
\acresetall
\section*{Supplemental Material}
\label{App:Supplemental analysis}

\subsection*{Simulation Techniques}
Both global and zoom-in simulations are performed using \texttt{AthenaK}~[37--39].
Our code solves the equations of ideal general-relativistic magnetohydrodynamics in dynamical spacetimes \cite{Banyuls:1997zz, Anton:2005gi} using a 2nd order finite-volume approach.
In particular, we use the 5th order \texttt{WENOZ} \cite{2008JCoPh.227.3191B} scheme to reconstruct primitive variables to cell interfaces and the Harten--Lax--van Leer--Einfeldt (HLLE) approximate Riemann solver to compute numerical fluxes \cite{Harten1983, Einfeldt1991}.
The magnetic field is evolved using the upwind constrained transport scheme of Gardiner \& Stone \cite{Gardiner:2005hy, Gardiner:2007nc, 2020ApJS..249....4S}, which preserves the no-magnetic-monopoles constraint to machine precision.
The background simulations evolve the spacetime metric using the Z4c formulation of Einstein equations \cite{Hilditch:2012fp}, discretized using a 6th order finite differencing scheme, as described in detail in Ref.~\cite{Daszuta:2021ecf} and Ref.~[38].
In the zoom-in simulations, we evaluate the spacetime metric by interpolating the data from the background simulation, as discussed in detail below.
We use a $3$rd order, strong-stability preserving Runge--Kutta scheme for the time integration~\cite{Gottlieb2001}.

Neutron star matter is modeled using the finite-temperature, composition-dependent, SFHo equation of state \cite{Steiner:2012rk}.
Accordingly, matter inside the neutron stars is assumed to be composed of nucleons, electrons, and photons.
Nuclear interactions are accounted for using a relativistic mean-field model.
Nuclear statistical equilibrium is assumed at low densities.
Our simulations do not explicitly include neutrino transport.
However, we add neutrinos to the equation of state following the approach described in \cite{Perego:2019adq}, which adequately captures the impact of
neutrinos in the high-density, high-temperature regime relevant for our simulations~\cite{Espino:2023dei}.

\subsection*{Initial and Boundary Conditions}

\subsubsection*{Global Simulations}

We perform a simulation of an equal-mass binary \ac{NS} merger.
The initial data is generated with the {\tt Lorene} pseudo-spectral code \cite{Gourgoulhon:2000nn}.
The initial configuration consists of two neutron stars with masses $m_1=m_2=1.3~M_\odot$ at a separation of $45~{\rm km}$.
The simulation domain consists of a box with sides of length $4535\ {\rm km}$, with $192$ points per dimension on the base level, plus $8$ additional nested refinement levels.
We set the innermost level covering a box of size $\approx (74,74,30)\ {\rm km}$ around the center of mass of the binary, so that the stars are completely contained during the full inspiral.
The intermediate levels are automatically defined by the code provided the above parameters and the number of points per meshblock (see Ref.~[37]), which we set to $32^3$.


On top of the hydrodynamic solution, we seed each of the neutron stars with a dipole-like field, defined through a vector potential which approximately describes the field generated by a current loop: 
\begin{equation}
    \mathcal{A}_\phi = \frac{\pi r_0^2 I_0 \varpi^2}{\left( r_0^2 + r^2 \right)^{3/2}} \left[ 1 + \frac{15r_0^2(r_0^2+\varpi^2)}{8\left(r_0^2+r^2\right)^2} \right],
\end{equation}
where $\varpi^2=(x-x_{\rm NS})^2+y^2$, $r^2=\varpi^2+z^2$, $I_0$ is the current, and $r_0$ is the radius of the current loop.
We choose $r_0\approx 3~{\rm km}$ and $I_0$ such that the maximum magnetic field, at the center of each star, is $B_{\rm max}\approx 2\times 10^{12}~{\rm G}$.
This choice provides a realistic magnetic field intensity at the surface of the stars \cite{Lorimer2008LRR....11....8L}
The magnetic field extends outside the stars and fills the whole simulation domain.

\begin{figure*}
    \centering
    \includegraphics[width=0.99\linewidth]{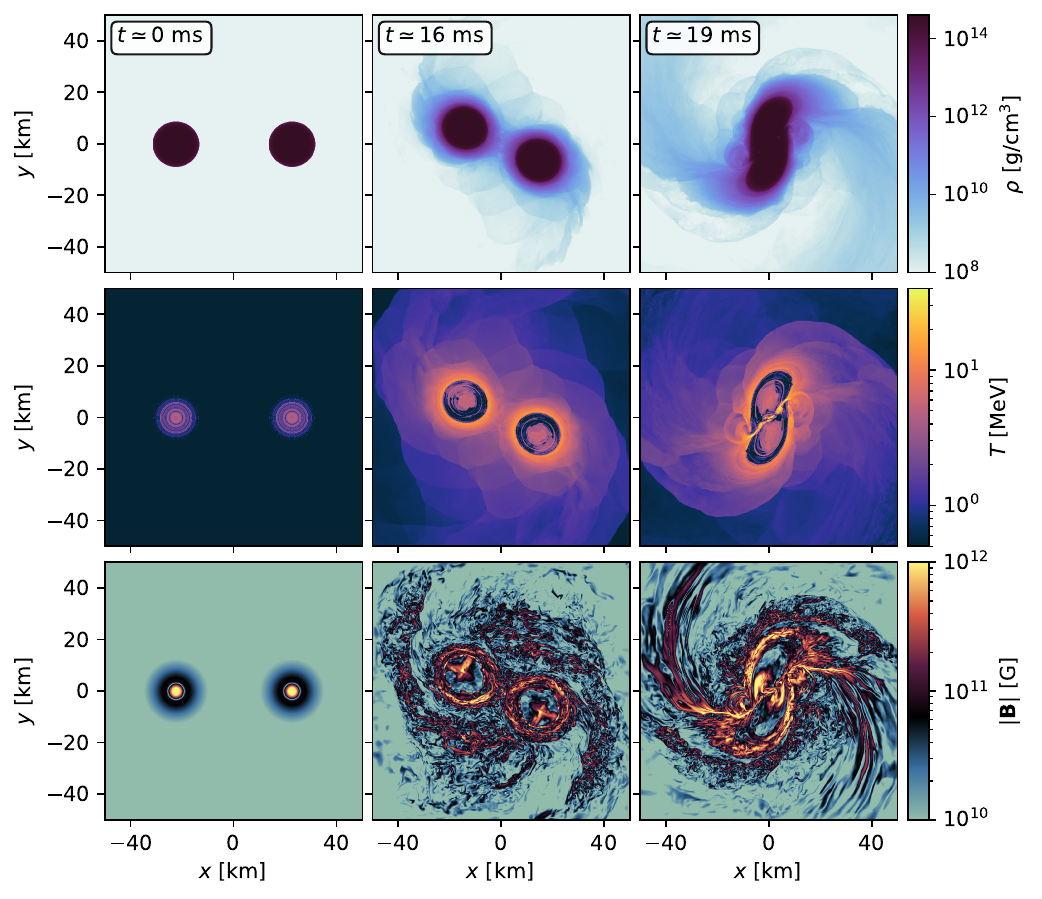}
    \caption{{\bf Equatorial slices of representative variables from the global simulation.} From top to bottom: Rest-mass density ($\rho$), temperature ($T$), and magnetic field strength($|{\bf B}|$). From left to right: $t\simeq 0, 16, 19$ ms.}
    \label{fig:global}
\end{figure*}

We evolve the binary for ${\sim} 7$ orbits until the stars merge, and follow the post-merger for approximately $5~{\rm ms}$, covering the initial magnetic field amplification driven by the \ac{KHI}.
Figure \ref{fig:global} shows equatorial slices of mass density ($\rho$, left), temperature ($T$, center), and vertical magnetic field ($B^z$, right) a three different times for the global simulations.

\subsubsection*{Metric and Fluid Data}

During the global simulation, we record simulation data in a cube of side length $2 L$, $L=4.5\ G M_\odot / c^2 \simeq 6.65\ {\rm km}$ centered about the origin.
This data is used to provide initial data, boundary conditions, and background metric for the zoom-in simulations.
To avoid possible boundary effects, the domain of the zoom-in simulations is restricted to a smaller cube with side length $2 L'$, $L' = 4\ G M_\odot / c^2 \simeq 5.91\ {\rm km}$.
The data is stored on a Chebyshev endpoint grid with $N = 64$ points in each direction and sampled with a frequency of $\Delta t = 0.09375\ G M_\odot / c^3 \simeq 0.46~\mu {\rm s}$.
In particular, we store and interpolate the fluid baryon number density $n_{\rm b}$, the Lorentz-factor-weighted three velocity $\tilde{v}^i = W v^i$, $W$ being the Lorentz factor, the pressure $p$, and the proton fraction $Y_{\rm e}$ of the plasma.
The spacetime geometry is encoded in the lapse and shift vectors $\alpha$ and $\beta^i$, respectively, the three metric $\gamma_{ik}$, and the extrinsic curvature, $K_{ik}$.
See, e.g., Ref.~\cite{2012LNP...846.....G} for a definition of these quantities.
Additionally, we store the Eulerian-frame densitized magnetic field $\mathcal{B}^i = \sqrt{\gamma} B^i$, $\gamma$ being the determinant of the three-metric.

We interpolate all fields, except for the magnetic field (which is described in detail below), using $6$th order Lagrange interpolation on the Chebyshev grid.
This results in a speedup of a factor ${\sim}(N/5)^3 \simeq 2\times 10^3$ compared to the direct evaluation of the Chebyshev collocation basis, which would have been prohibitively expensive, as we need to interpolate spacetime data and boundary conditions at each timestep of the zoom-in simulation.
We have verified that the error introduced by this truncation of the interpolation stencil is negligible, as can be expected since the weight associated with each collocation basis function decreases exponentially with distance from the collocation point for smooth data.
This is the same approach used by \texttt{SpEC} to couple spacetime evolution, performed on spectral grids, and matter evolution, performed on a uniformly spaced grid \cite{Duez:2008rb}.

\begin{figure}
    \centering
    \includegraphics[width=0.994\linewidth]{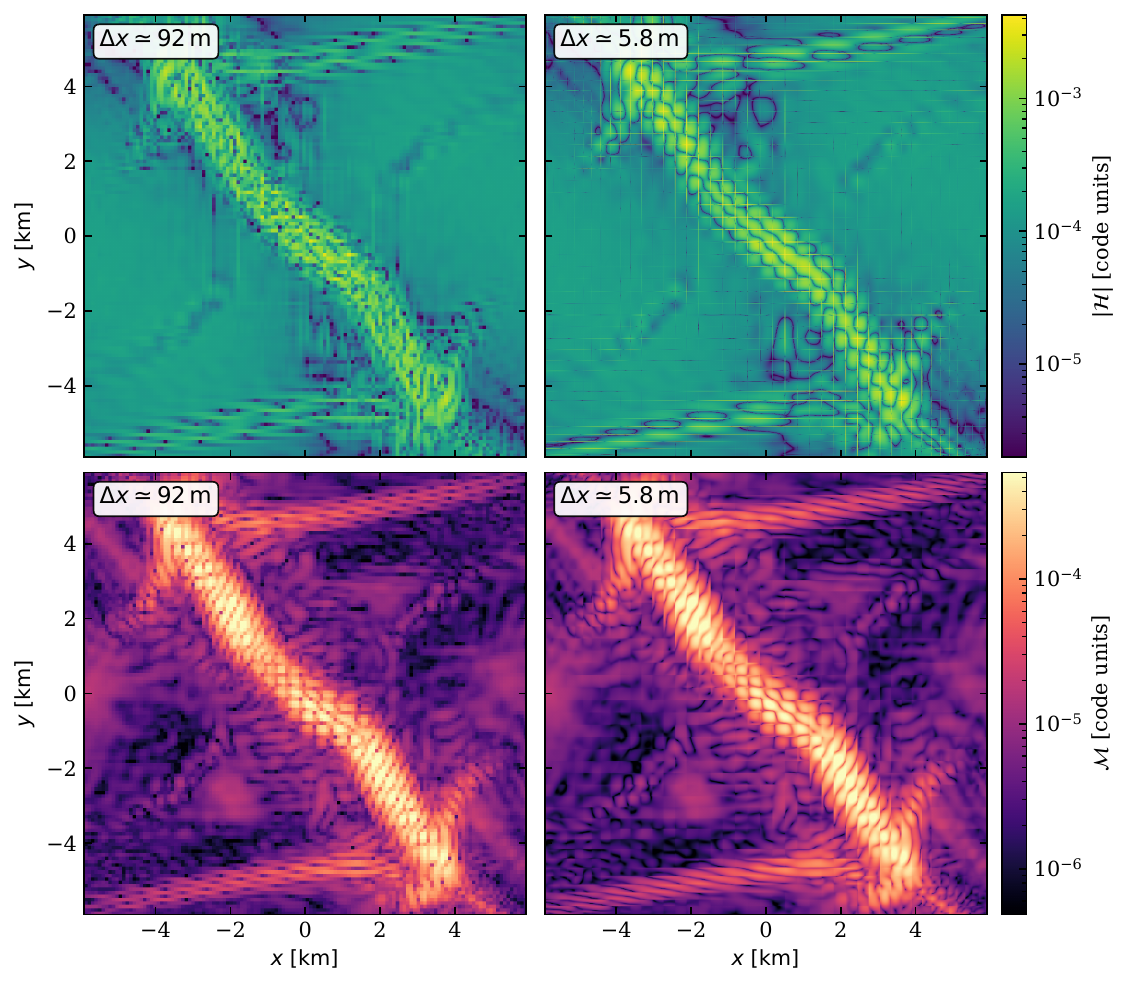}
    \caption{Equatorial slices of the Hamiltonian constraint and the norm of the momentum constraint in the zoom-in domain.
    Cell-centered $xy$ slices near the equatorial plane of the absolute Hamiltonian residual $|\mathcal H|$ (top row) and the momentum-constraint diagnostic $\mathcal M$ (bottom row) for the $\Delta x\simeq 92~{\rm m}$ (left) and the $\Delta x\simeq 5.8~{\rm m}$ (right) simulations.}
    \label{fig:constraints}
\end{figure}

To characterize local violations of the Einstein constraints introduced by the mapping, we show in Fig.~\ref{fig:constraints} equatorial maps of the Hamiltonian
constraint (top row) and the norm of the momentum (bottom row) constraint, $M:= \sqrt{M^i M_i}$ at the initial time $t \simeq 0$, for the coarsest and for the finest unigrid zoom-in simulations, $\Delta x \simeq 92~{\rm m}$ (left column) and $\Delta x \simeq 5.8~{\rm m}$ (right column).
These are computed using the prescribed metric together with the matter fields evolved in the zoom-in simulation.
The residuals are of order $10^{-5}-10^{-4}$ over most of the domain and reach at most $10^{-3}$, with the largest values confined to the shear layer, where the gradients of the matter fields are steepest.
We note that these maps include the error incurred in interpolating the matter fields, which enter the constraints through the energy density and momentum
density and which are considerably less smooth than the metric; they are therefore an upper bound on the violation attributable to the mapping of the spacetime alone, and the concentration of the largest values along the shear layer reflects this.
Comparing the two columns, the residuals are almost unchanged between the two resolutions, despite the grid spacing differing by a factor of $16$.
We also note that these constraints are consistency conditions for a self-gravitating spacetime-and-matter solution; exact satisfaction with the evolved fluid is not required to solve the GRMHD equations on a prescribed spacetime.
The accuracy of the fluid evolution instead depends on the prescribed metric and its derivatives, whose numerical errors are part of the background approximation.

\subsubsection*{Magnetic Field Interpolation}

The interpolation of the magnetic field requires additional considerations to avoid the creation of magnetic monopoles.
\texttt{AthenaK} applies boundary conditions using the standard ``ghost cell'' approach, whereby the domain is artificially extended by a few grid points on which data is prescribed.
In the case of the zoom-in simulation, we prescribe data in this region by interpolation from the global simulation.
To this end, it is possible to directly interpolate the magnetic field.
The reason is that, in the context of the upwind constrained transport, the ghost data is only used to compute the electric field at the cell edges.
The magnetic field, which is collocated on the cell faces, is then updated using Faraday's law, so its time derivative is divergence-free by construction.
In other words, no divergence constraint violation can flow into the domain from the boundaries.

The initial conditions are more delicate, because the constraint-transport algorithm guarantees that the divergence of the magnetic field should stay constant to machine precision. Still, unphysical artifacts can be expected if the initial magnetic field is not solenoidal.
As a first step, we reconstruct the vector potential using the identity \cite{2010JGRA..11510112W}
\begin{equation} \label{eq:homotopy}
    \mathcal{A}_i(x) = \int_0^1 \epsilon_{ijk}\,
    \mathcal{B}^j(\lambda x)\, x^k\, \lambda\, \mathrm{d}\lambda\, ,
\end{equation}
where $\epsilon_{ijk}$ is the antisymmetric symbol with $\epsilon_{123} = 1$.
The integral in Eq.~\eqref{eq:homotopy} is computed in pre-processing for each grid point using a Gaussian quadrature with $12$ points.
To this aim, the $\mathcal{B}$ field data is interpolated to the quadrature points using a $10$th order accurate Lagrange interpolation.

Next, we interpolate the vector potential $\mathcal{A}_i$ at the cell edges of the zoom-in grid, using the same $6$th order interpolation approach as for the other
variables, and evaluate
\begin{equation}
    \mathcal{B}^i = \epsilon^{ijk}\, \partial_j\, \mathcal{A}_k
\end{equation}
using the discrete curl operator consistent with the constrained transport of \texttt{AthenaK}, e.g.,
\begin{multline}
    \mathcal{B}^1_{k,j,i+1/2} = \frac{(\mathcal{A}_3)_{k,j+1/2,i+1/2} - (\mathcal{A}_3)_{k,j-1/2,i+1/2}}{\Delta x^2} -  \\
    \frac{(\mathcal{A}_2)_{k+1/2,j,i+1/2}-(\mathcal{A}_2)_{k-1/2,j,i+1/2}}{\Delta x^3}\,,
\end{multline}
where we have used the notation $f_{k,j,i}$ to denote the field at location $x^i = x_0^i+ (i\, \Delta x^1, j\, \Delta x^2, k\, \Delta x^3)$.

\subsection*{Calculation of the Power Spectra}

We compute power spectra for the density weighted three-velocity $\mathbf{w}=\sqrt{\rho} \mathbf{v}$ and for the Eulerian-frame magnetic field $\mathbf{B}$
as
\begin{equation}
    P_v(\mathbf{k}) = \frac{1}{2}\, \hat{\mathbf{w}}^\ast \cdot \mathbf{w}, \qquad
    P_B(\mathbf{k}) = \hat{\mathbf{B}}^\ast \cdot \mathbf{B},
\end{equation}
where $\hat{\cdot}$ denotes the 3D Fourier transform and $\cdot^\ast$ complex conjugation.
Since we are not working in a periodic domain, we window the data using a symmetric Tukey window before the Fourier transformation.
The Fourier transforms are computed in post-processing using the FFTW Python bindings.

The metric fields are smooth across the computational domain, with relative spatial gradients of representative quantities such as the lapse $\alpha$ and $\sqrt{\gamma}$ satisfying $|\nabla X|/X \lesssim 0.08~\mathrm{km}^{-1}$.
At this level, deviations from a locally flat spacetime are small, and the spectra can be meaningfully interpreted in analogy with flat-space scalings such as the Kolmogorov and Kazantsev scalings.

One-dimensional spectra are defined as
\begin{equation}
    P_v(k) = \int P_v(\mathbf{k})\, \delta(|\mathbf{k}|-k)\, \mathrm{d}^3\mathbf{k},
\end{equation}
where $\delta$ is the Dirac delta.
The one-dimensional magnetic field spectrum is defined in an analogous way.
Note that, since we are working in a finite domain, the spectra are quantized, so all integrals reduce to summations.
Integrals over spherical shells are transformed to weighted sums following
\cite{1988CF.....16..257E}:
\begin{equation}
  P_v(k) = \frac{4\pi k^2}{N_k} \sum_{k-1/2 < |\mathbf{k}| \leq k+1/2} P_v(\mathbf{k})\,,
\end{equation}
where $N_k$ is the number of discrete wave-numbers in the shell $k-1/2 < |\mathbf{k}| \leq k+1/2$.

\subsection*{Magnetic field evolution}

\begin{figure*}
    \centering
    \includegraphics[width=0.995\linewidth]{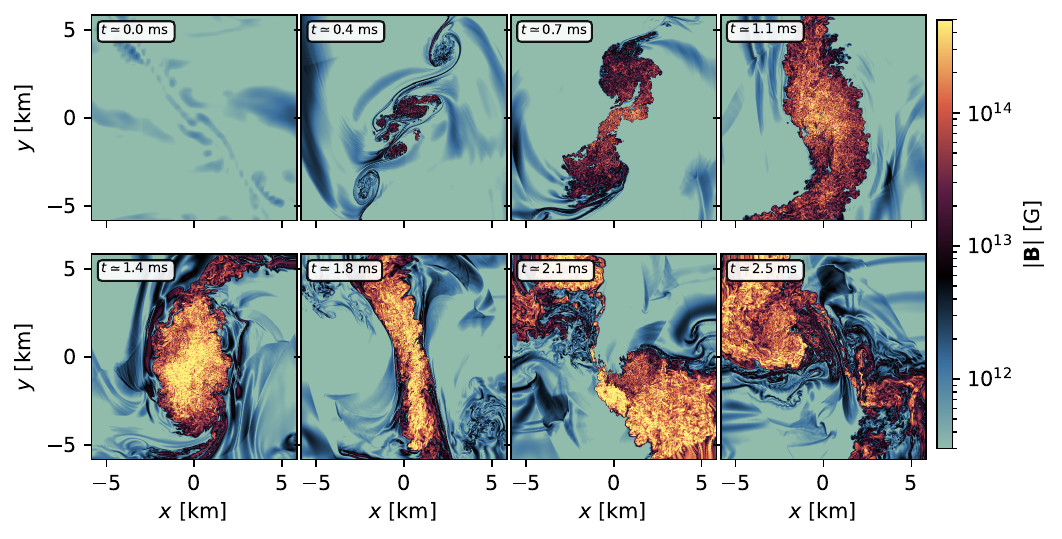}
    \caption{{\bf Evolution of the shear layer and magnetic field.}
Equatorial slices of the magnetic-field strength at selected times for the $\Delta x \simeq 2.9~{\rm m}$ simulation, showing the development of Kelvin--Helmholtz-driven turbulence.}
    \label{fig:2D_B_evolution}
\end{figure*}

In this Section, we provide additional diagnostics and analysis of the magnetic field evolution in our simulations.
The exponential amplification of the field seen in our simulations is consistent with the action of a small-scale dynamo driven by turbulence.
Similar amplification mechanisms operate in other turbulent astrophysical environments such as the interstellar medium (e.g., Beattie et al. \cite{2025MNRAS.542.2669B, 2025NatAs...9.1195B}).
In Figure \ref{fig:2D_B_evolution}, we show 2D snapshots of the magnetic field evolution at several times from the beginning of the simulation until the end.
The initially smooth shear layer rapidly develops vortical structures due to the \ac{KHI}.
Several large vortices form along the layer, which in turn become themselves KH-unstable and become turbulent approximately after one turnover time; the magnetic field lines are stretched, twisted, and folded within these regions and are exponentially amplified.
As time passes, the initial vortices merge into larger ones and ultimately coalesce into a single, thick turbulent layer.
The magnetic field continues to be amplified at small scales due to the turbulence until approximately $2$ ms.
The process continues with the turbulent layer separating again into 2 or more coherent structures with large magnetic fields due to the nonlinear global dynamics.

\begin{figure*}
    \centering
    \includegraphics[width=0.9\linewidth]{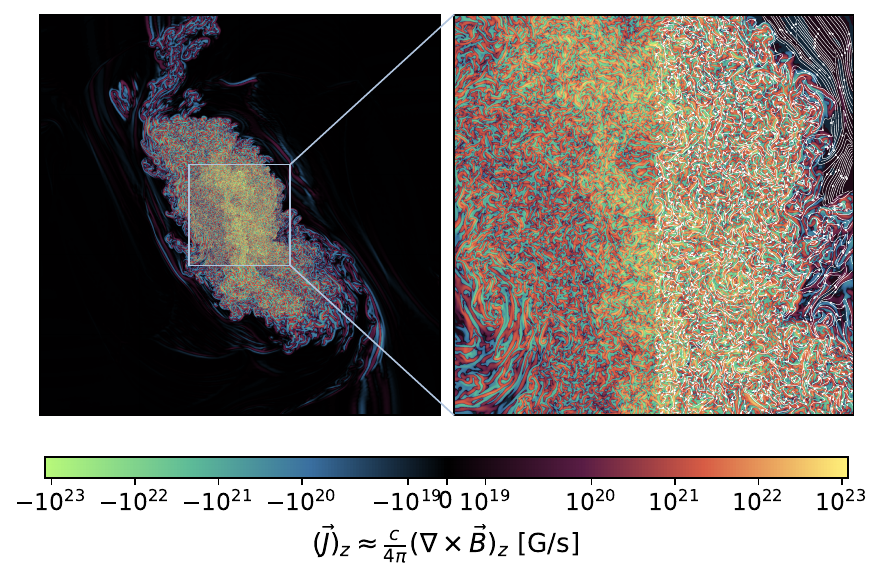}
    \caption{{\bf Intermittent current-sheet formation.}
    Equatorial snapshot of the vertical component of the (Newtonian) current density $\textbf{j}=(c/4\pi)~\nabla \times \textbf{B}$ at $t\simeq 1.5~{\rm ms}$.
    The right panel shows a zoomed-in region of size ${\approx} 3$ km; magnetic field lines are superimposed on the right half of the plot, tracing the intermittent current sheet structures.}
    \label{fig:curlB}
\end{figure*}

Turbulence driven by the \ac{KHI} leads to the formation of small, intermittent current sheets embedded within the turbulent cascade.
These structures arise from the accumulation of vortices across a range of scales and the onset of secondary instabilities.
Because the current density $\textbf{j}\approx(c/4\pi)~\nabla \times \textbf{B}$ directly traces magnetic-field gradients, it provides a particularly clear diagnostic of regions where the field is strongly stretched, twisted, and folded by the flow, and where magnetic tension, pressure forces, and reconnection become dynamically important.
Regions of enhanced current, therefore, highlight locations where small-scale kinetic energy is efficiently transferred into magnetic energy.
In Fig.~\ref{fig:curlB}, we show the component of the current density perpendicular to the equatorial plane at $t\simeq 1.5~{\rm ms}$; it demonstrates how the small-scale current sheets permeate the full volume of the turbulent region.

\begin{figure*}
    \centering
    \includegraphics[width=0.9\linewidth]{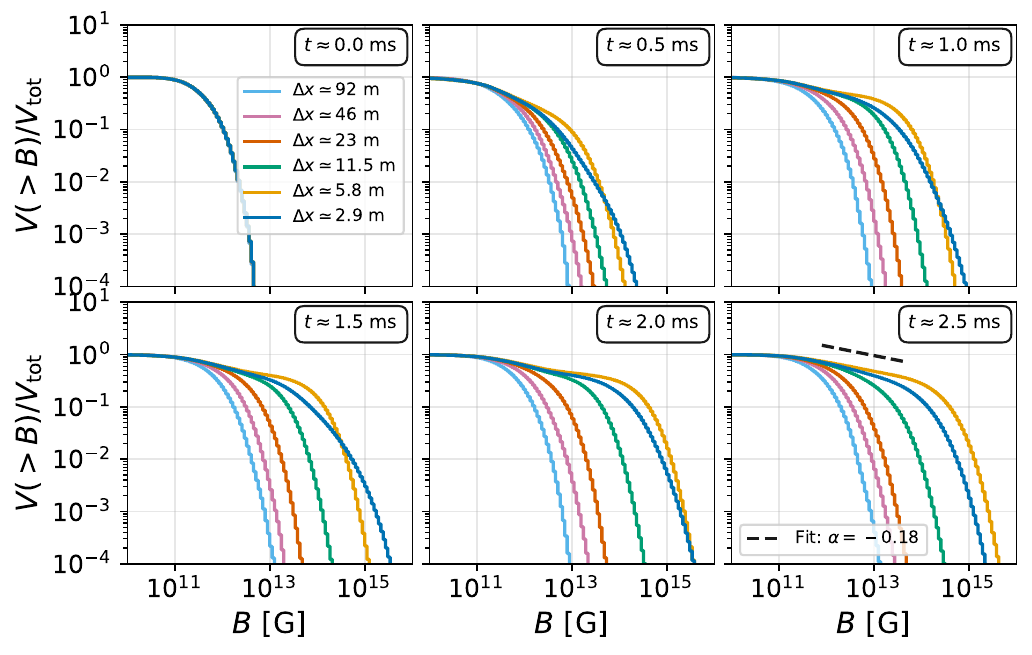}
    \caption{{\bf Cumulative volume distribution of magnetic-field strength.}
Cumulative volume fraction, $p(B>B_0)$, at different times for all simulations.
In the final panel, a power-law fit (short dashed line) is shown for the unigrid $\Delta x \simeq 5.8~{\rm m}$ simulation.}
    \label{fig:histograms}
\end{figure*}

The intermittency of magnetic-field amplification in small-scale turbulent eddies can be quantified using the cumulative volume distribution for the magnetic field strength, $p(B>B_0)$, defined as the fraction of the volume within which the magnetic field exceeds a certain value $B_0$.
In other words, this measures the propensity of a given region of the plasma to have a magnetic field amplified to a given strength.
In Figure \ref{fig:histograms}, we show the time evolution of this quantity for all our simulations at different numerical resolutions.
At early times, the magnetic field is weak throughout most of the volume, with the bulk of the domain having $B \lesssim 10^{11}~{\rm G}$ and only a small portion reaching field strengths above $10^{12}~{\rm G}$.
Once the magnetic amplification sets in, all simulations rapidly develop exponential tails with high-$B$ in the cumulative distribution.
As expected, the extent of this tail is resolution-dependent: higher-resolution runs populate more regions with progressively higher magnetic-field strengths.
By $t\sim 1$ ms, the two highest-resolution simulations already exhibit a non-negligible portion of the domain with fields $>5\times 10^{14}$ G.

The shape of the exponential tail depends on the simulation setup.
In particular, our highest-resolution run employs a mesh-refinement strategy where only a fraction of the volume attains the maximum resolution, while a substantial portion has approximately half the resolution of the unigrid simulation with $N=2048^3$ ($\Delta x \simeq 5.8\ {\rm m}$).
As a result, at times $>1.5~{\rm ms}$, this run exhibits a shallower tail, with fewer regions at $B\sim 10^{14}~{\rm G}$, but extending to higher field strengths than the unigrid $N=2048^3$ case.
By the end of the runs, both high-resolution simulations have developed a distinct power-law tail at magnetic-field strengths below those of the exponential tail; this feature is not clearly visible in the lower-resolution runs.
From $t\simeq 2$ ms until the end of the simulations, the cumulative distribution follows a power-law $p(B>B_0)\sim B_0^{-p}$, with $p\approx 0.18$.
The emergence of such power-law tails is consistent with strongly intermittent magnetic-field amplification driven by the \ac{KHI}-induced turbulence.

\begin{figure}
    \centering
    \includegraphics[width=0.95\linewidth]{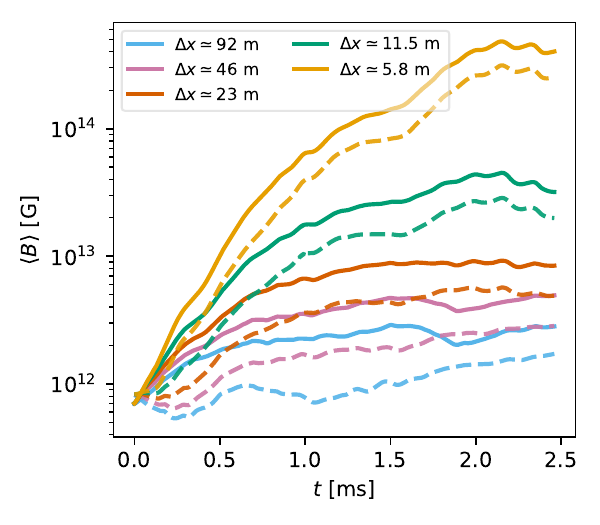}
    \caption{
{\bf Time evolution of the average toroidal (solid lines) and poloidal (dashed lines) magnetic field strength for our simulations.}
    }
    \label{fig:pol_tor}
\end{figure}

\begin{figure*}
    \centering
    \includegraphics[width=0.5\linewidth]{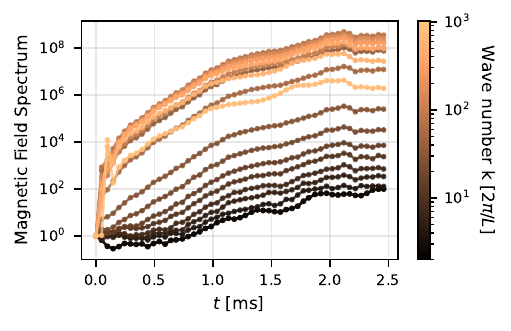}
    \includegraphics[width=0.46\linewidth]{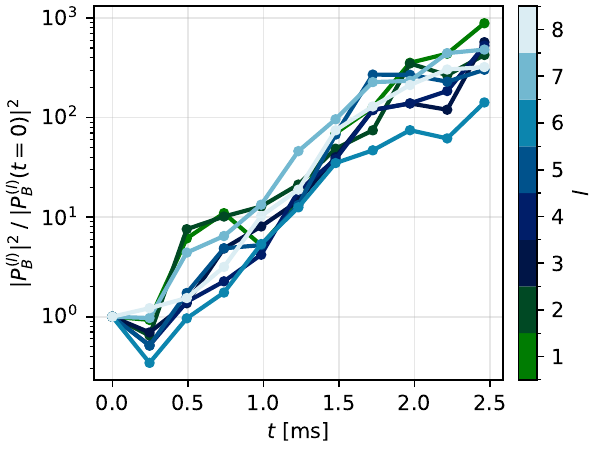}
    \caption{{\bf Time evolution of the magnetic-field Fourier and spherical modes.}
    {\it Left:} Magnetic-field power as a function of time for individual wavenumbers $k$ in the $\Delta x \simeq 5.8 ~ {\rm m}$ simulation.
    {\it Right:} Time evolution of the magnetic multipoles for the $\Delta x \simeq 5.8 ~ {\rm m}$ simulation.
    }
    \label{fig:spectra_time}
\end{figure*}

In Figure \ref{fig:spectra_time}, we show the time evolution of the magnetic-field power spectrum, resolved into individual Fourier modes $k=0,...,1024$.
Early in the run, magnetic energy grows exponentially at high wavenumbers, corresponding to small spatial scales, while power at larger scales remains nearly unchanged.
This behavior reflects the initial amplification of the field by small-scale turbulence associated with the \ac{KHI}.
As the simulation proceeds, power at progressively lower wavenumbers begins to increase, after the small-scale field has reached sufficiently large amplitudes.

In Figure \ref{fig:pol_tor}, we show the time evolution of the average magnetic field, separated into its poloidal and toroidal components for our simulations.
At the highest resolutions, both components grow exponentially while maintaining an approximately constant ratio.
In particular, the growth of the toroidal component far exceeds what differential rotation alone can produce over the few rotation periods covered by our simulations, since winding supplies only a linear growth of the field.
This behavior is consistent with the turbulent dynamo amplifying both components at a similar rate.

\begin{figure}
    \centering
    \includegraphics[width=0.95\linewidth]{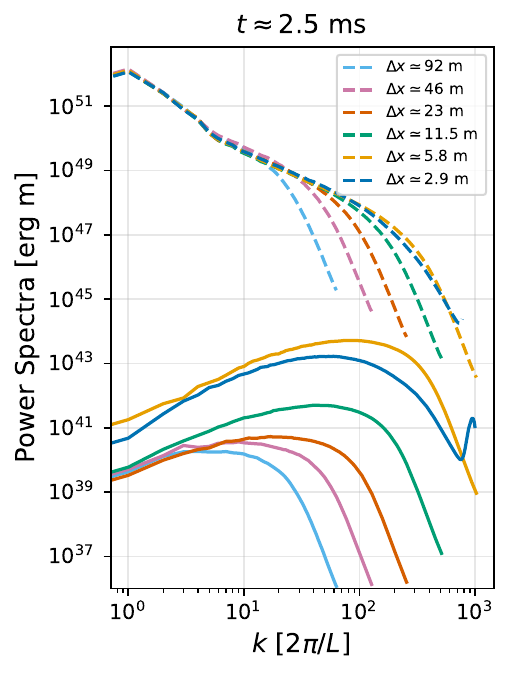}
    \caption{
{\bf Kinetic and magnetic power spectra at different numerical resolutions.}
Dashed lines show kinetic spectra; solid lines show magnetic spectra.
}
    \label{fig:spectra_together}
\end{figure}

Finally, in Figure \ref{fig:spectra_together} we show the kinetic and magnetic energy spectra at the final time for all simulations.
The magnetic field does not reach equipartition with the kinetic energy, staying subdominant by approximately three orders of magnitude and maintaining the Kazantsev scaling.
 This contrasts with the monotonically decreasing spectrum in Ref.~[29] of the main manuscript; in our case, the field is dynamically weak and evolves in the kinematic regime, producing the characteristic rising spectrum of the small-scale dynamo regime.

An additional diagnostic to describe the topology of the magnetic field can be made by analyzing its (vector) spherical harmonic decomposition.
An orthonormal basis for such a decomposition at fixed radius $r$ is given by \cite{Barrera_1985}
\begin{equation}
    {\bf Y}_{lm}= Y_{lm} \hat{\bf r}, ~~~~~ {\bf \Psi}:=r{\bf \nabla} Y_{lm}, ~~~~~ {\bf \Phi}_{lm}:= \hat{\bf r} \times {\bf \Psi},
\end{equation}
where $\hat{\bf r}$ is the unit vector along the radial direction in spherical coordinates and $Y_{lm}(\theta,\phi)$ are the scalar spherical harmonics.
The three basis vectors ${\bf Y}_{lm}$, $ {\bf \Psi}$, and ${\bf \Phi}_{lm}$ correspond, respectively, to the purely radial (longitudinal), poloidal, and toroidal components of the field.

The spherical multipole moments of the magnetic field are then obtained by projecting ${\bf B}$ onto each basis vector and integrating over the full solid angle at fixed radius $r$.
To characterize the global contribution of each multipole, we further integrate these coefficients over radius (weighted by $r^ 2$); then, we calculate the multipole modes as
\begin{align}
     B_{lm}^{(r)}=\int_V {\bf B}\cdot {\bf Y}_{lm}^{*} \sqrt{\gamma}dV, \\
     B_{lm}^{(1)}=\frac{1}{l(l+1)}\int_V {\bf B}\cdot {\bf \Psi}_{lm}^{*} \sqrt{\gamma}dV, \\
     B_{lm}^{(2)}=\frac{1}{l(l+1)}\int_V {\bf B}\cdot {\bf \Phi}_{lm}^{*} \sqrt{\gamma}dV,
\end{align}
where $\sqrt{\gamma}$ is the square root of the spatial metric determinant, and the factor $1/l(l+1)$ ensures proper normalization.
To quantify the power in each multipole, we define
\begin{equation}
    P_B^{(l)}:= \sum_{X=\{r, (1), (2)\}}~\sum_{m=-l}^l |B_{lm}^{(X)}|^2
\end{equation}
The $l=1$ mode corresponds to the dipolar component of the field, while higher $l$ modes capture progressively smaller-scale structure, analogous to higher wavenumbers $k$ in the Fourier decomposition.
The right panel of Fig. \ref{fig:spectra_time} shows the time evolution of the power in each multipole up to $l=8$, normalized to its initial value.
All multipoles exhibit substantial amplification.
The dipolar component ($l=1$) grows at a rate comparable to the higher-order modes, reaching amplitudes up to three orders of magnitude larger than its initial value.
This behavior is consistent with the turbulent dynamo operating over a broad range of angular scales.

\end{document}